# Ordering kinetic in two-dimensional hexagonal pattern of cylinder-forming PS-*b*-PMMA block copolymer thin films: dependence on the segregation strength

Gabriele Seguini,[1,*] Fabio Zanenga,[1,2] Michele Laus,[2] Michele Perego.[1,*]

[1] *Laboratorio MDM, IMM-CNR, Via C. Olivetti 2, I-20864 Agrate Brianza, Italy.*
[2] *Dipartimento di Scienze e Innovazione Tecnologica (DISIT), Università del Piemonte Orientale ''A. Avogadro'', INSTM, UdR Alessandria, Viale T. Michel 11, I-15121 Alessandria, Italy.*

\* (GS) gabriele.seguini@mdm.imm.cnr.it   \* (MP) michele.perego@mdm.imm.cnr.it

**ABSTRACT**

This paper reports the experimental determination of the growth exponents and activation enthalpies for the ordering process of standing cylinder-forming all-organic polystyrene-*block*-poly (methyl methacrylate) (PS-*b*-PMMA) block copolymer (BCP) thin films as a function of the BCP degree of polymerization (N). The maximum growth exponent of 1/3 is observed for the BCP with the lowest N at the border of the order disorder transition. Both the growth exponents and the activation enthalpies exponentially decrease with the BCP segregation strength ($\chi \cdot N$) following the same path of the diffusivity.

**TEXT**

A two dimensional (2D) hexagonal lattice (domains with crystalline structure) represents the equilibrium, low temperature, low symmetry phase of various spatiality modulated phase systems, i.e. crystalline solids, colloidal systems, and block copolymer (BCP) thin films. [1–4] The size and the number of atoms or colloids are fixed, conversely, in BCPs the size of the domains can deform and their number can change over time. The equilibrium state corresponds to the absolute minimum free energy of the system. The similarities extend to the types of topological defects, disclinations, dislocations, and grain boundaries. [5–7] The Kosterlitz-Thouless-Halperin-Nelson-Young theory describes the 2D melting for a system in complete thermodynamic equilibrium by means of a two-step continuous transition. [8–10] Experimentally, the polycrystalline pattern coarsens in time toward the equilibrium state by annihilation of the topological defects reducing the grain boundary length and by reorientation of ordered domains through thermally activated kinetic processes. The grain size of the hexagonal lattice growths in time (t) scaling as $t^\phi$, where $\phi$ is the growth exponent. [11,12] The ordering process is relevant both fundamentally to understand the far from equilibrium coarsening phenomena and practically because the order affects the functional properties of the systems.

In BCP self-assembly, the nanostructures emerging from the phase separation are encoded in the molecular composition of the constituent polymer chains. BCPs consisting of two covalently end joined polymeric blocks are single component systems as the chemically different blocks cannot macro phase separate. [13] The covalent bond between the two blocks constrains the system to micro phase segregate. The chain composition (f) determines primarily the equilibrium morphology of the ordered phase, moving from lamellae to cylinders, gyroids, spheres increasing the disproportion between the two chains. [14] A 2D hexagonal structure results from a single layer matrix of spherical domains or from thin films comprising domains of cylinders perpendicularly oriented with respect to the substrate. [15] The degree of polymerization (N), i.e. the number of monomeric units in the polymer chain, sets the characteristic dimensions, diameter (d) of the domains and periodicity ($L_0$) between the domains. The Flory–Huggins parameter ($\chi$) quantifies the thermodynamic interaction between the two distinct monomeric units and accounts for the change in local free energy. The BCP free energy, that determines the equilibrium state of the system, is composed by an entropic contribution ($\Delta S \div 1/N$) and by an enthalpic one ($\Delta H \div \chi$). The overall segregation strength is hence governed by $\chi \cdot N$. With positive $\chi$, the competition between $\Delta H$ reduction, because of the local segregation between the different blocks, is counterbalanced by the $\Delta S$ decrement due to the localization of the interface between the blocks and to the stretching of the chains. When $\chi \cdot N \ll 10$, the dominant $\Delta S$ drives the system in a spatially homogenous phase. At $\chi \cdot N \approx 10$, the two opposite effects are balanced defining the order disorder transition (ODT) boundary in the phase diagram. For $\chi \cdot N \gg 10$, $\Delta H$ dominates and microphase segregation appears. [13] According to this picture, $\chi \cdot N$ establishes the minimum attainable size of the domains in the phase separated system. [16] The thermodynamic driving force for the phase separation increases with N or, when $\chi$ is temperature (T) dependent, decreasing T. However, for high N, or low T, the ordering kinetic is limited by the reduced diffusivity of the system. The real ordering process results from the balance between the slow kinetic and the enhanced thermodynamic. [8]

Due to their potential applications as advanced tool for nanofabrication, accurate and reliable approaches have been developed to control BCP self-assembly. [17–19] This broad experimental knowledge makes BCPs appealing as model systems for uncovering the interplay between thermodynamic driving forces and kinetic barriers during the ordering process. Mesoscopic order results from the competition between a segmental short-range attractive and a chain long-range repulsive interaction. [6,12,20,21] Experimentally, BCP thin films are prepared by a spin coating procedure, where solvent mediates the unfavorable interactions between the blocks. Solvent evaporation prevents the motion of the chains and the resulting spin-coated polymer film is frozen in a metastable state. [22,23] An anneling step is necessary to release this phase. The temperature processing window for microphase ordering is comprised between the glass transition temperature ($T_G$) and the onset of thermal degradation of the polymer chains. [24] Thermal annealing (TA), [25] increasing T, or solvent vapor annealing (SVA), [23,26] adding a solvent, facilitate the phase separation and the defect annihilation through the increased chain mobility. [26] TA and SVA processes are thermodynamically equivalent as they can be mapped on the BCP phase diagram considering the solvent effect as a kinetic (plasticization, reduction of $T_G$) or as a thermodynamic effect (dilution, reduction of $\chi$). [14] A rapid TA (RTA) approach provides the possibility to reach target T in very short time, even though not instantaneous, taking advantage of the residual solvent trapped in polymer film to boost the chain mobility and speed-up the ordering process. [27,28] This solvent-assisted RTA (SARTA) treatment was demonstrated to be a valuable tool to probe the ordering process in phase separated BCP thin films. [27–32] In more details, the RTA treatment is composed of three parts, a heating ramp, an isothermal step at constant $T_A$, and a cooling ramp. For heating rates equal or higher than 18 °C/s, the final level of order of the system is mainly determined by the isothermal step, with negligible contribution of the heating ramp. On the other hand the cooling step is fixed by the machine recovery time. [32]

This letter aims to experimentally determine the ordering kinetic dependence on $\chi \cdot N$ in 2D hexagonal pattern of strongly segregated cylinder-forming polystyrene-*block*-poly (methyl methacrylate) (PS-*b*-PMMA) BCP thin films. To promote perpendicular orientation of the domains in the PS-*b*-PMMA BCP thin films, poly (styrene-*random*-methyl methacrylate) (P(S-*r*-MMA)) random copolymer (RCP) (molecular weight, $M_n$ = 69 kg/mol, Styrene fraction, $f_s$ = 0.61, polydispersity index, PDI = 1.19) was grafted (RTA: T = 310 °C, t = 60 s) on hydroxyl terminated Si surface to form a neutral ≈ 19 nm thick brush layer. This thick RCP brush layer provides a reservoir of embedded solvent [29,30,33] able to sustain the ordering process in the phase separated PS-*b*-PMMA film over several decades of time during the SARTA treatment. [29] The all-organic PS-*b*-PMMA BCP exhibits $\chi_{S-MMA}$ parameter that is weakly dependent on T, because the entropic contribution (0.021) is much larger than the enthalpic one (3.2/T). [34–36] This enables studies of the ordering process over a wide range of T keeping χ almost constant and modulating χ·N, i.e. the driving force for the phase separation, by changing N. [34,37] Asymmetric PS-*b*-PMMA BCPs with $f_S \approx 0.7$, PDI ≤ 1.11, and different N ranging from 461 to 1281 (Table 1) were spinned in toluene solution to obtain ≈ 30 nm thick layer. The ordering process in the BCP thin films was systematically studied considering different combinations of annealing time ($t_A$) and temperature ($T_A$). For each $T_A$, typically comprised between 160 and 290 °C, grain coarsening evolution was assessed at $t_A$ = 1, 9, 90, and 900 s. Slightly different $T_A$ ranges were defined for each BCP depending on N. [31] Heating rate was fixed at 18°C/s for all the samples. Irrespective of the annealing parameters, increasing N, d varies from 12.5 to 28.6 nm while $L_0$ changes from 26.5 to 59.0 nm following the law $L_0 \div N^{2/3}$. [31] The 2/3 exponent sets the investigated BCPs in the strong segregation limit (SSL), thus resulting in sharp interfaces between the segregated blocks. [38] For N < 461, χ·N approaches the critical value corresponding to ODT, preventing the possibility to investigate ordering process over a proper range of ($T_A$, $t_A$) values. [24]

Surface morphology of the self-assembled BCPs was investigated by scanning electron microscopy (SEM) images. [31] Because of the probe depth of electrons, the top-down SEM gives real-space information of the BCP domains only near the free surface. Considering the thickness of the BCP film, the cylindrical domains are expected to propagate throughout the entire film depth. [31] Representative high-magnification SEM images at each ($T_A$, $t_A$) combination for different BCPs are reported in (SM, 1-6). The order of the 2D hexagonal pattern was quantified by measuring the correlation length (ξ) using Delaunay triangulation to detect the pattern orientation in SEM images, [14] providing information over an area that is much larger than the measured correlation length, as previously detailed in ref. [29]. This procedure rules out possible finite size effects, within the limits of this real space method. [14] When $\xi > L_0$ ($1 < \xi < 2L_0$ and $\xi > 2L_0$ correspond to blue and green SEM images, respectively, in SM, 1-6) the system, after the (rapid) early-stage phase separation, enters the (slow) late-stage coarsening regime. [39] In this regime, the average grain dimensions regularly increase their order as a function of the increased $T_A$ or $t_A$ for all the investigated N. At higher $T_A$ and $t_A$, inhomogeneities begin to appear on the sample surface (red SEM images in SM, 1-6). These samples were not considered for the quantitative evaluation of $\xi$ evolution. For each ($T_A$, $t_A$) combination, average $\xi$ values were obtained from the analysis of at least five SEM images acquired in different regions of the sample. The dispersion of the data was evaluated as the standard deviation from the average. The measured $\xi$ value can be treated following the time-Temperature superposition (tTS) procedure, to describe the ordering kinetics. [29,32] $\xi$ as a function of $t_A$ for different $T_A$ are collected in a master curve at the reference $T_A$ (Table 1) versus an equivalent time ($t_{EQ}$) spanning over several decades. All the master curves can be fitted with a simple power law, irrespective of N (SM, 7-12). This regular behavior confirms the pertinence of the tTS procedure and evidences a self-similar time evolution.

Figure 1 reports the evolution of $\xi/L_0$ as a function of $t_{EQ}$ for different BCPs. All the ordering curves are normalized by dividing the grain dimension ($\xi$) by the inherent periodicity of the specific BCP ($L_0$). This evidences the number of ordered rows of the cylinder domains irrespective of N, i.e. of the physical dimensions of the domains. [31] $\xi/L_0$ curves follow a power law $t^\phi$, with different $\phi$ values (from $\approx 1/3$ to $\approx 1/10$) depending on N (Table 1). Interestingly $\phi$ is not constant within the cylindrical morphology but it is N-dependent, that is, basically, it depends on the driving force for the phase separation. [40] $\phi$ results 1/3 for N = 523 and no further increase is observed decreasing N, suggesting 1/3 as the maximum $\phi$ for strongly segregated standing cylinder domains resulting from BCP self-assembly. Increasing N, progressive reduction of ϕ is observed. Interestingly, a ϕ ≈ 1/3 value equals the growth exponent for systems with only attractive interaction that coarsen with a rate limiting step that is a diffusion limited mechanism. [20,41] Literature data for thermally treated standing cylinders in PS-*b*-PMMA thin films report 1/4 for BCP with N = 652, consistently with our experimental results. [42,43]

Figure 2 reports ϕ values as a function of $N_{MMA}$, i.e. the length of the smaller of the two blocks. For $N_{MMA}$ > 167, ϕ decreases as a function of $N_{MMA}$, following an exponential decay:

$$\phi \sim e^{-\chi \cdot N_{MMA}} \qquad (1)$$

where χ has a value of 0.028 ± 0.002, in excellent agreement with $\chi_{S-MMA}$ values reported in the literature. [34,37] This result indicates that the grain coarsening process is retarded with a barrier proportional to the segregation strength $\chi \cdot N_{MMA}$. Note that the exponent, which depends on both χ and N, in this specific BCP system is almost T independent. ϕ value for $N_{MMA}$ =125 is much lower than expected according to the extrapolation of Eq (1), suggesting that the exponential dependence on the thermodynamic variable $\chi \cdot N_{MMA}$ holds only in the SSL, that is for $\chi \cdot N_{MMA} = O(\chi \cdot N_{MMA}^{ODT})$, where monomeric unit segregation impedes chain diffusion across different grains. Chain diffusion in phase separated BCP follows the same exponential decay of ϕ as a function of $\chi \cdot N_{MMA}$, because of the thermodynamic penalty for the diffusion due to the increased localization of the chain with N. [15,44–46] Ohonogi et al. [11] reported simulations showing that ϕ depends on the value of the noise strength. It becomes progressively larger increasing noise, moving from a value ≈ 0.15 at zero noise up to a limiting value just above ≈ 0.30. The noise strength corresponds to a diffusion at the microscopic scale [47] and it is proportional to T. [6] Sagui et al. [20] evaluated the growth exponent for a system with competing interactions phase separated into the exagonal phase founding a value of 0.29. The exponent decreases to 0.18 increasing the long range repulsion interaction. [20,21] Ashwani et al. [40]

reported a similar slow down of the coarsening for stripe patterns increasing the distance from the transition point. These literature results are perfectly consistent with data herein presented, considering the decrease of diffusivity occurring because of the progressive N increase. Besides, Adland et al. [2] noted that for continuous lattice pattern of spatially modulated phase such as BCPs, because the polymer chains can stretch changing the shape of the domains, [8] the growth is limited by dissipation in the grain interior associated with lattice translation. Their simulations predict ϕ value of 0.22 considering bulk dissipation. Differently, increasing T, i.e. increasing chain diffusivity, the reduction of the number of defects by dislocation reaction allows the grain to shrink with less rotation, reducing the contribution of the grain interior dissipation. ϕ value of 0.35 is foreseen in case of minimized grain interior dissipation, consistently with our experimental findings for small N values.

To elucidate the kinetics of the grain coarsening process, the evolution of $\ln(\xi)$ as a function of $1/T_A$ at different $t_A$ has been investigated (SM, 13-18). Experimental data can be fitted using an Arrhenius equation $\ln \xi = \ln A - H_A / (K_B \cdot T_A)$, where $H_A$ corresponds the activation enthalpy of an elementary step of the grain coarsening process, $K_B$ is the Boltzmann constant, and A is the pre-exponential factor. The linear evolution of $\ln(\xi)$ as a function of $1/T_A$ indicates that the grain coarsening process is a thermally activated process with a kinetic barrier. Limited variations of $H_A$ with $t_A$ have been observed for each N, resulting in a mean $H_A$ value with limited dispersion for each specific BCP (Table 1). Figure 3 reports the mean $H_A$ values as a function of $N_{MMA}$. For $N_{MMA} > 167$, $H_A$ data follow an exponential decay as a function of $N_{MMA}$:

$$H_A \sim e^{-\chi \cdot N_{MMA}} \qquad (2)$$

where χ has a value of 0.024 ± 0.005, consistently with the χ value obtained by fitting ϕ data using eq. (1) and data in the literature. [34,37] It is worth to note that, for grain coarsening of standing hexagonal cylinder, Majewski et al. [48] reported a value of 69 ± 23 kJ/mol for a 49 kg/mol (N ≈ 461) BCP, and Ji et al. [42] reported 24-43 kJ/mol for a 67 kg/mol (N = 523) BCP, in perfect agreement with our results.

Present $H_A$ values are much lower than free energy barriers associated to topological defect creation. [49–51] Actually, topological defects in self-assembled BCP template results from a collection of many molecules and they cannot be considered as equilibrium fluctuations around a perfectly ordered state. On the contrary, they correspond to non-equilibrium metastable states, kinetically frozen during phase separation and the initial stages of the self-assembling process. [49–51] The excess free energy (penalty) to create a defect is large and increases with χ·N. The corresponding equilibrium defect density is small and exponentially decreases with this penalty. [37,49–51] The free energy barriers for defect removal is much lower than the excess free energy to create a defect. These free energy barriers are of the same order of magnitude of the $H_A$ values we measured for the grain coarsening process on unpatterned flat surfaces, but their evolution as function of N follows an opposite trend compared to $H_A$. [1,50]

The decrease in the grain coarsening kinetic and of the activation enthalpy with increasing N can be rationalized in the framework of the ubiquitous thermodynamic compensation effect. [52] The inset of Figure 3 reports the relationship between $H_A$ and ln(A) i.e. the pre-exponent of the Arrhenius equation, that is related to the activation entropy $S_A$. The linear relationship indicates that these data follow the Meyer-Neldel (MN) rule [53,54] i.e. $\Delta S = \Delta H/T_{MN}$, where $T_{MN}$ is the MN temperature. In thermally activated processes, going from the metastable ground state to the activated one, the number of paths to the activated state supplied by the heat bath increases with $H_A$, resulting in an increase of entropy for the system. For $T > T_{MN}$ an increase of $H_A$ corresponds to a large positive change $\Delta S$, lowering the free-energy barrier that limits grain coarsening. From another point of view, it has been shown that the compensation is a consequence of a balance between repulsive and attractive interactions. [55] From the slope of the linear fitting of the data in the inset of Figure 3, $T_{MN}$ is 428 K (155 °C). Considering that for these BCPs the reported mean bulky $T_G$ of PS is $\approx$ 378 K (105 °C) and of PMMA is $\approx$ 397 K (124 °C) it results that $T_{MN} \approx 1.1\ T_G$. Above $T_G$ and below the scales of the domain size, blocks can be considered as liquid. [15] In this context, a value above $\approx 1.2 \cdot T_G$ sets the threshold where the inverse relationship between diffusivity and viscosity holds. [56–59]

In conclusion, reported data demonstrate connections between the BCP equilibrium state and the ordering kinetic of the system. In this framework, a model system (PS-*b*-PMMA) along with a suitable process (SARTA) emerge as a platform for fundamental studies where a collection of various universal behavior can be simultaneously scrutinized. The small dependence on T of $\chi_{S-MMA}$ allows following the ordering process as a function of N, indeed using 1/N instead of T as a unique variable to modify the thermodynamic driving force that guides the system toward the equilibrium state and investigate the thermally activated kinetic of pattern formation in BCP thin films. The molecular architecture of the BCP encodes the characteristic dimensions (d, $L_0$) as well as the activation enthalpy ($H_A$) and the kinetic ($\phi$) of the ordering process. Furthermore, the growth exponent data highlights two facts; the exponential decay with $N_{MMA}$ and a maximum limiting value of 1/3 for BCP in the SSL. The former mimics the decrease of the diffusivity with $N_{MMA}$ highlighting that the grain coarsening process is limited by single chain diffusion. The latter sheds light to the mechanism of ordering process in the general contest of the rate of decrease of the total excess grain boundary free energy in a polycrystal. Moreover, the linear dependence between $H_A$ and $S_A$ sets the system in the ubiquitous thermodynamic compensation rules where the linear factor is correlated to the $T_G$ of the BCP and it is the transition temperature to the activated coarsening state.

## ACKNOWLEDGMENTS

The authors acknowledge K. Sparnacci, V. Gianotti, D. Antonioli (Università Del Piemonte Orientale, Italy) for their assistance in RCP synthesis, M. Alia (CNR-IMM, Italy) for assistance in the

experimental part, T. J. Giammaria (CNR-IMM, Italy) and F. Ferrarese Lupi (INRIM, Italy) for fruitful discussion. This research has been partially supported by the project "IONS4SET" funded from the European Union's Horizon 2020 research and innovation program under grant agreement No. 688072.

**TABLE**

*Table 1 Molecular Weight (M$_n$), Polydispersity Index (PDI), degree of polymerization (N), Styrene fraction (f$_S$), MMA degree of polymerization (N$_{MMA}$), Annealing temperatures (T$_A$), reference T$_A$ for tTS, diameter (d), periodicity (L$_0$), growth exponent ($\phi$), activation enthalpy (H$_A$) for the different BCPs.*

| M$_n$ (Kg/mol) | PDI | N | f$_S$ | N$_{MMA}$ | T$_A$ (°C) | T$_A$ (tTS) (°C) | d (nm) | L$_0$ (nm) | $\phi$ | H$_A$ (KJ/mol) |
|---|---|---|---|---|---|---|---|---|---|---|
| 47.5 | 1.07 | 461 | 0.74 | 125 | 160-240 | 180 | 12.5±0.5 | 26.5±0.5 | 0.32±0.02 | 65 ± 5 |
| 53.8 | 1.07 | 523 | 0.69 | 168 | 160-240 | 180 | 13.0±0.1 | 28.8±0.5 | 0.34±0.01 | 63 ± 5 |
| 67.1 | 1.09 | 652 | 0.69 | 210 | 190-270 | 190 | 17.0±0.1 | 35.0±1.0 | 0.17±0.01 | 42 ± 3 |
| 82.0 | 1.07 | 797 | 0.69 | 250 | 190-290 | 190 | 19.0±0.2 | 42.9±0.7 | 0.13±0.01 | 32 ± 2 |
| 101.5 | 1.08 | 988 | 0.67 | 335 | 190-290 | 190 | 22.7±1.5 | 47.0±1.0 | 0.10±0.01 | 28 ± 1 |
| 132.0 | 1.11 | 1281 | 0.73 | 355 | 190-270 | 190 | 28.6±1.6 | 59.0±4.0 | 0.08±0.01 | 26 ± 1 |

**FIGURES**

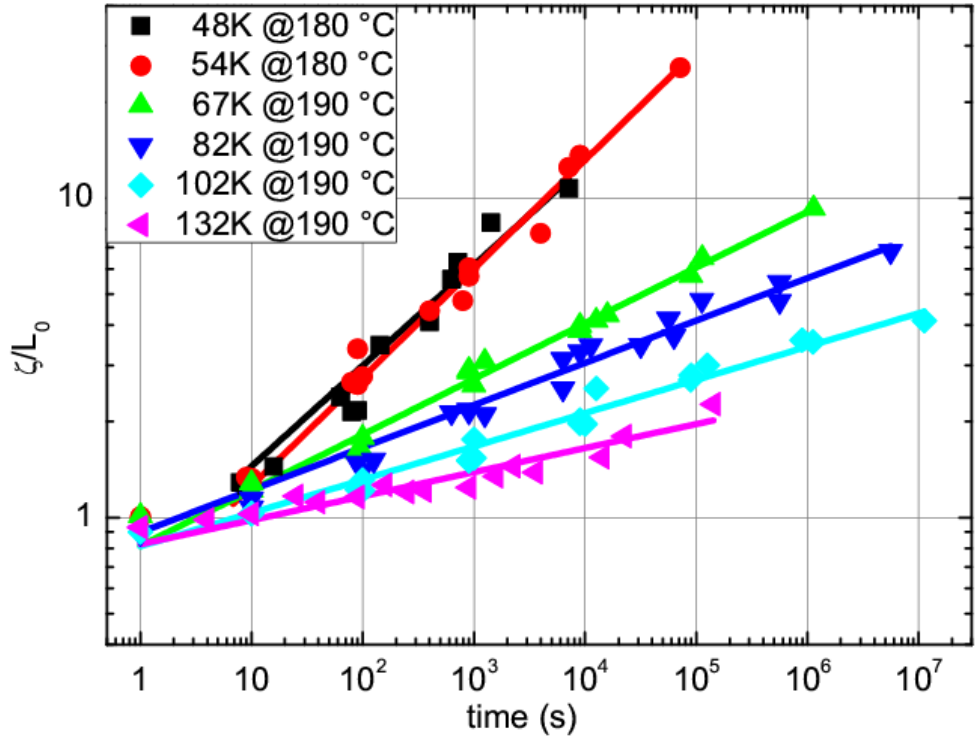

*Figure 1* Evolution of the correlation length normalized over the periodicity (ξ/L₀) as a function of the $t_{EQ}$ after the tTS procedure for BCP of different N (461, 523, 652, 797, 987, and 1281). The slopes of the curved represent the growth exponent φ.

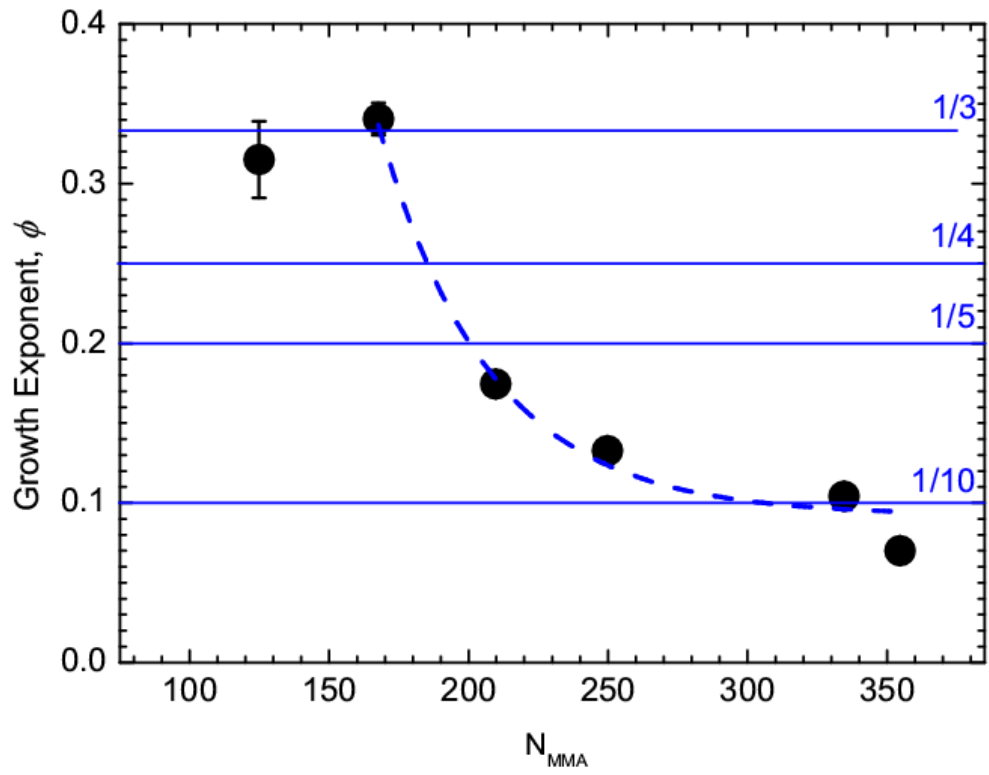

*Figure 2* Growth exponent $\phi$ as a function of $N_{MMA}$.

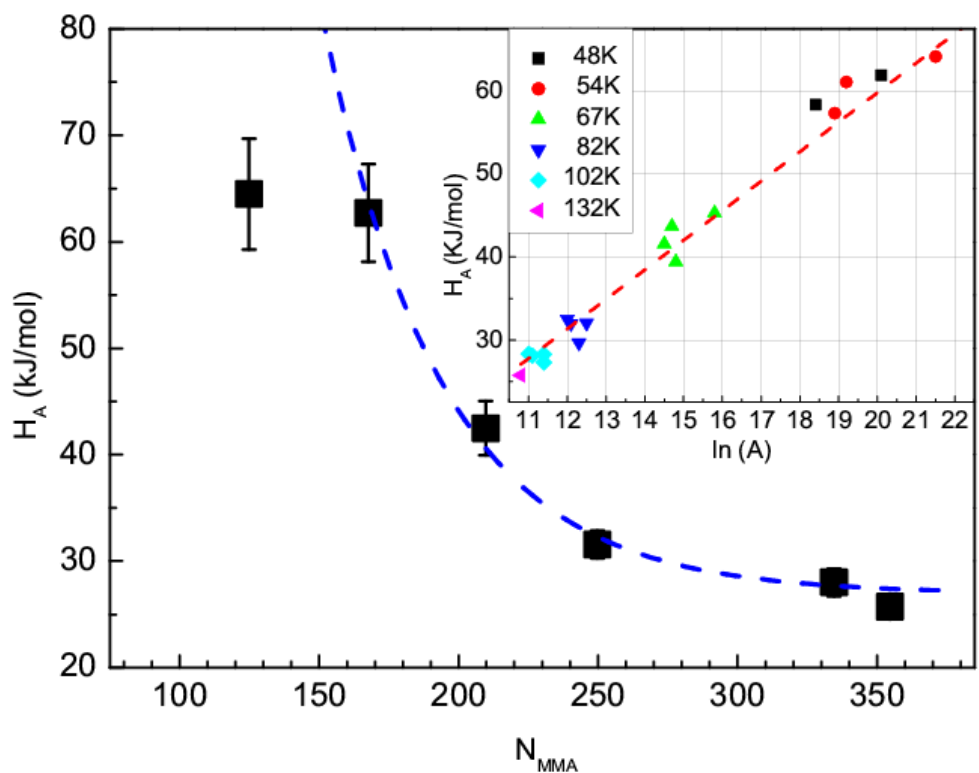

*Figure 3* Activation enthalpy $H_A$ as a function of $N_{MMA}$. (Inset) $H_A$ as a function of the pre-exponential factor (lnA) of the Arrhenius plot corresponding to the activation entropy ($S_A$).

# Ordering kinetic in two-dimensional hexagonal pattern of cylinder-forming PS-*b*-PMMA block copolymer thin films: dependence on the segregation strength

Gabriele Seguini,[1,*] Fabio Zanenga,[1,2] Michele Laus,[2] Michele Perego.[1,*]

[1] *Laboratorio MDM, IMM-CNR, Via C. Olivetti 2, I-20864 Agrate Brianza, Italy.*
[2] *Dipartimento di Scienze e Innovazione Tecnologica (DISIT), Università del Piemonte Orientale ''A. Avogadro'', INSTM, UdR Alessandria, Viale T. Michel 11, I-15121 Alessandria, Italy.*

\* (GS) gabriele.seguini@mdm.imm.cnr.it  \* (MP) michele.perego@mdm.imm.cnr.it

# Supplemental Material (SM)

## S1:      N = 461, $M_n$ = 48 kg/mol

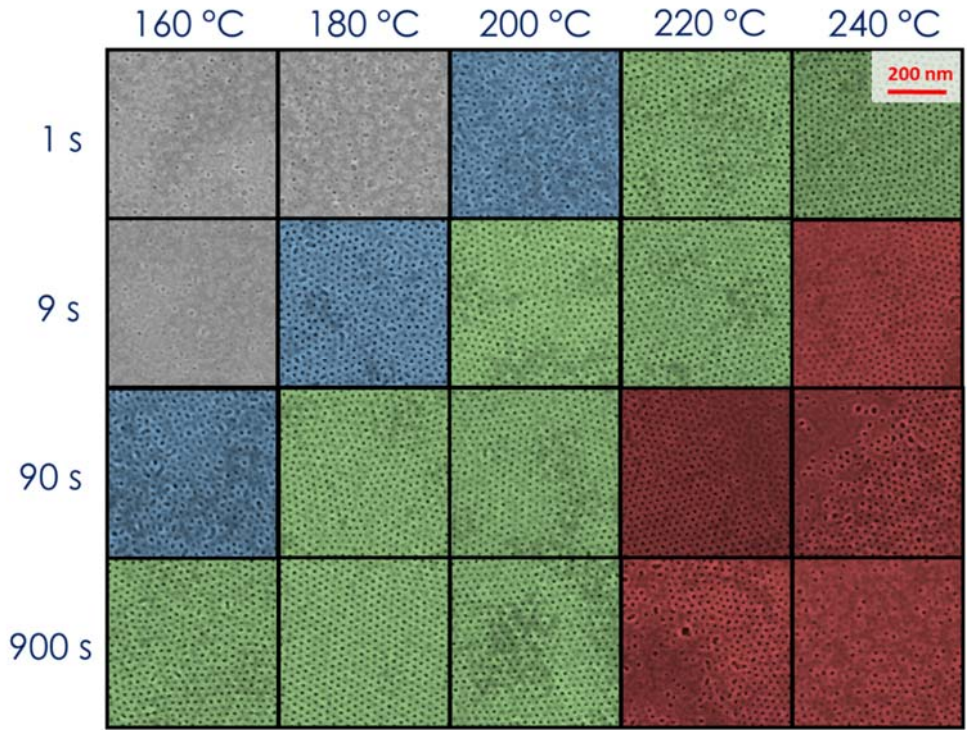

Figure 1 Plan view high-magnification SEM images of 48 kg/mol BCP thin films annealed at different temperatures $T_A$ for different $t_A$. $\xi < L_0$ gray images, $1 < \xi < 2 L_0$ blue images, $\xi > 2L_0$ green images, inhomogeneities red images.

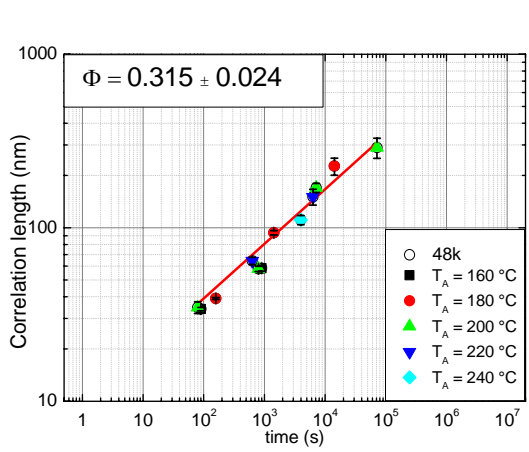

Figure 7 Master curve from time–Temperature-Superposition of the correlation length data

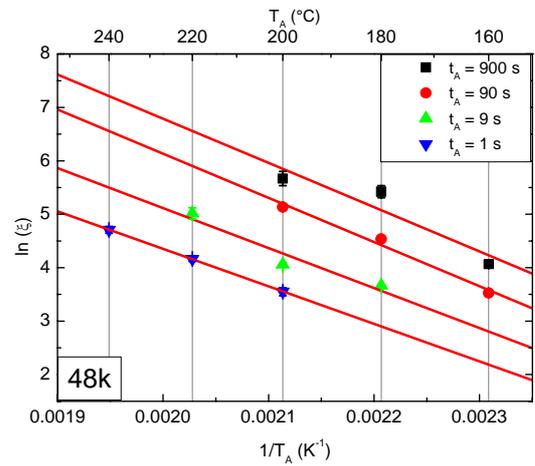

Figure 13 Arrhenius plot of $\xi$ data at different $t_A$.

S1

## S2:      N = 523, $M_n$ = 54 kg/mol

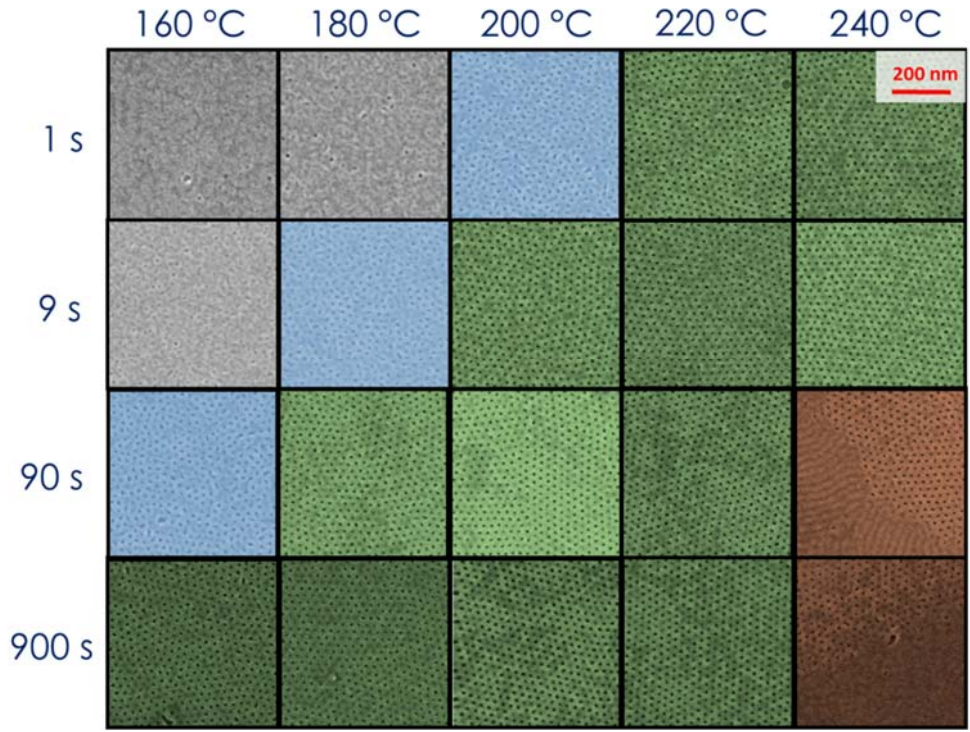

Figure 2 Plan view high-magnification SEM images of 54 kg/mol BCP thin films annealed at different temperatures $T_A$ for different $t_A$. $\xi < L_0$ gray images, $1 < \xi < 2 L_0$ blue images, $\xi > 2L_0$ green images, inhomogeneities red images.

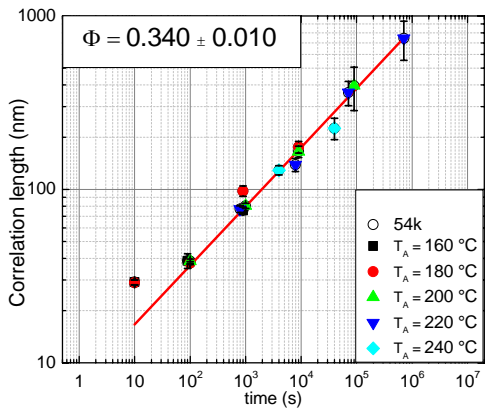

Figure 8 Master curve from time–Temperature-Superposition of the correlation length data

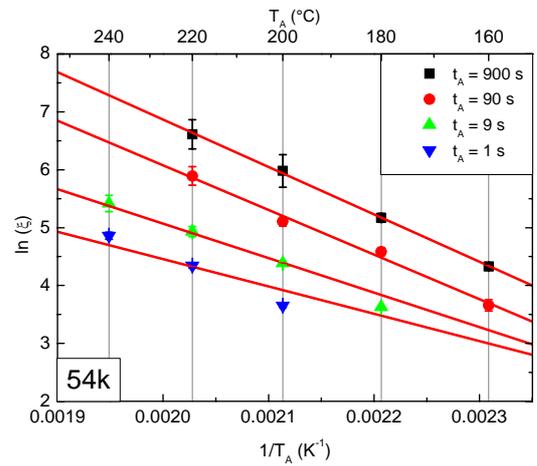

Figure 14 Arrhenius plot of $\xi$ data at different $t_A$.

S2

## S3:      N = 652, $M_n$ = 67 kg/mol

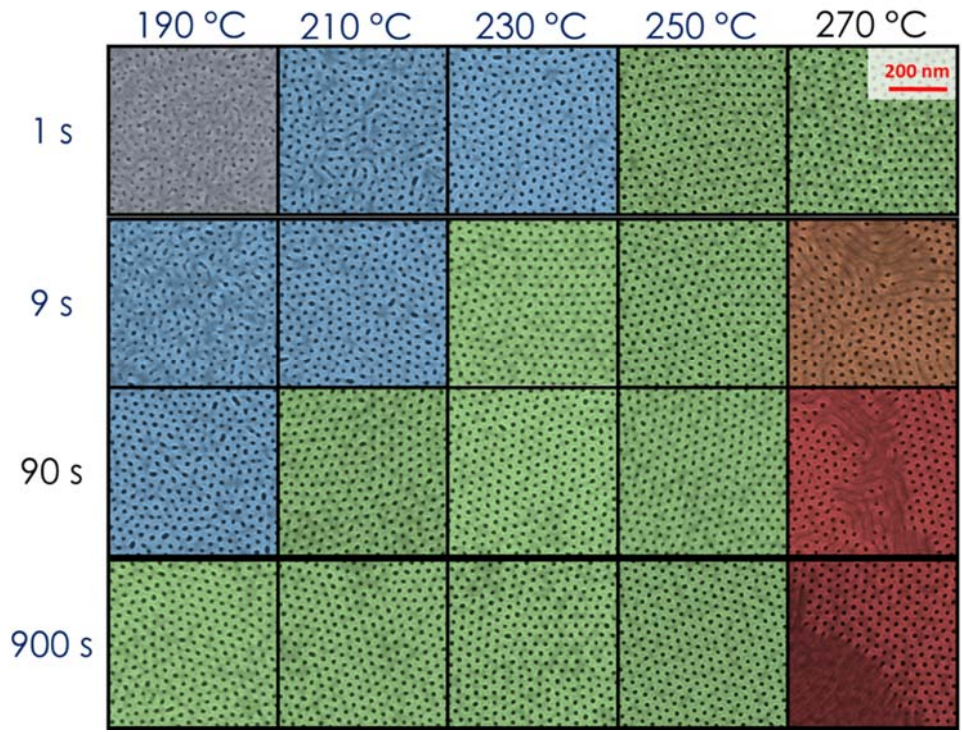

Figure 3 Plan view high-magnification SEM images of 67 kg/mol BCP thin films annealed at different temperatures $T_A$ for different $t_A$. $\xi < L_0$ gray images, $1 < \xi < 2 L_0$ blue images, $\xi > 2L_0$ green images, inhomogeneities red images.

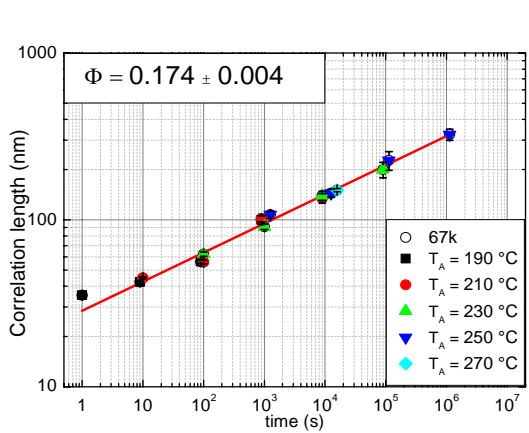

Figure 9 Master curve from time–Temperature-Superposition of the correlation length data

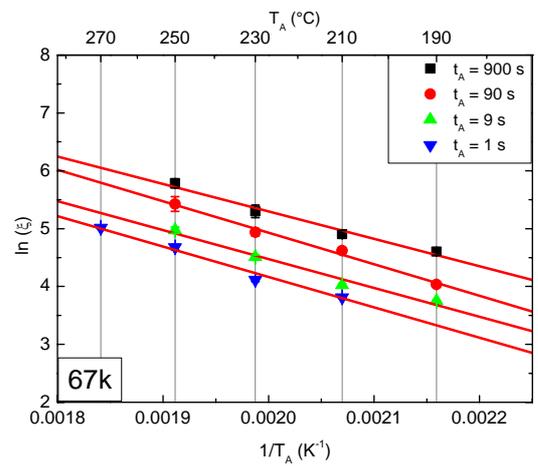

Figure 15 Arrhenius plot of $\xi$ data at different $t_A$.

# S4:         N = 797, $M_n$ = 82 kg/mol

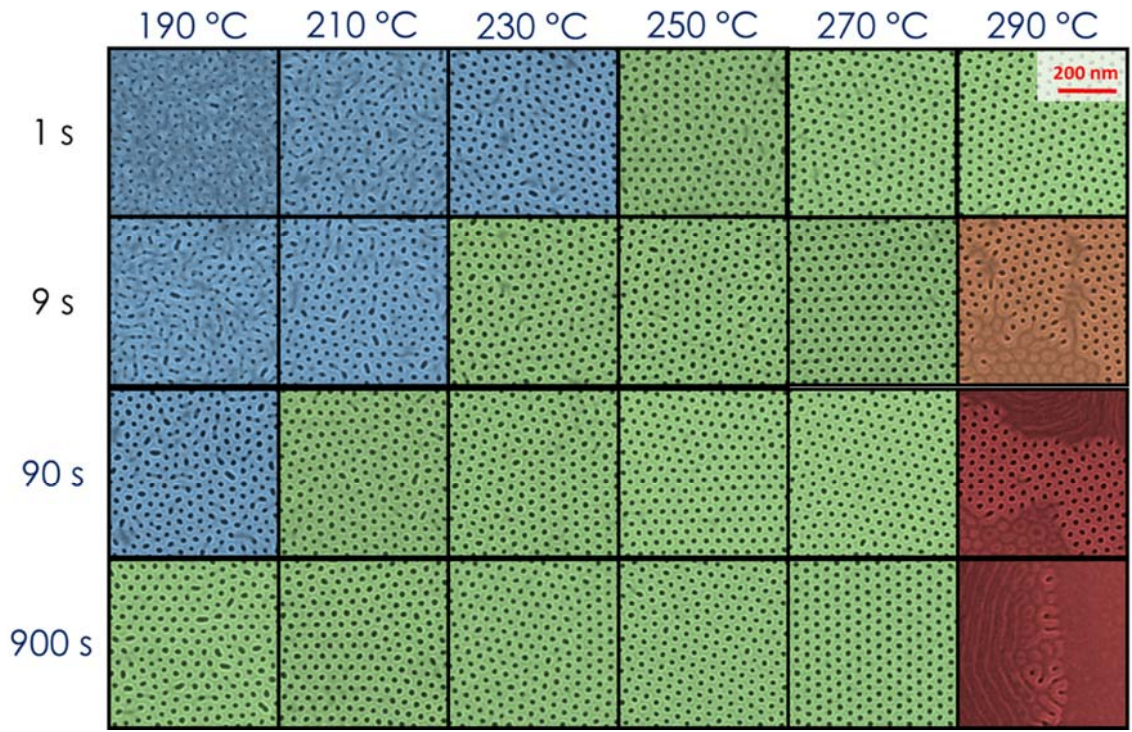

Figure 4 Plan view high-magnification SEM images of 82 kg/mol BCP thin films annealed at different temperatures $T_A$ for different $t_A$. $\xi < L_0$ gray images, $1 < \xi < 2 L_0$ blue images, $\xi > 2L_0$ green images, inhomogeneities red images.

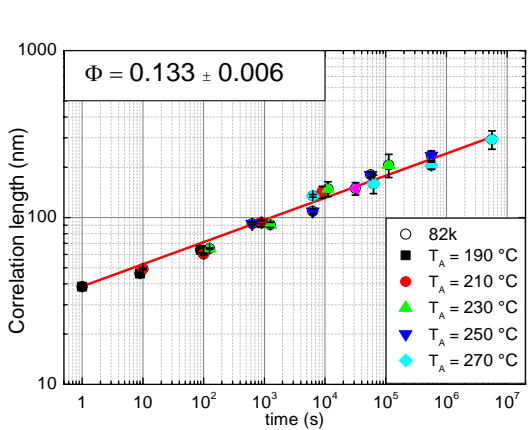

Figure 10 Master curve from time–Temperature-Superposition of the correlation length data

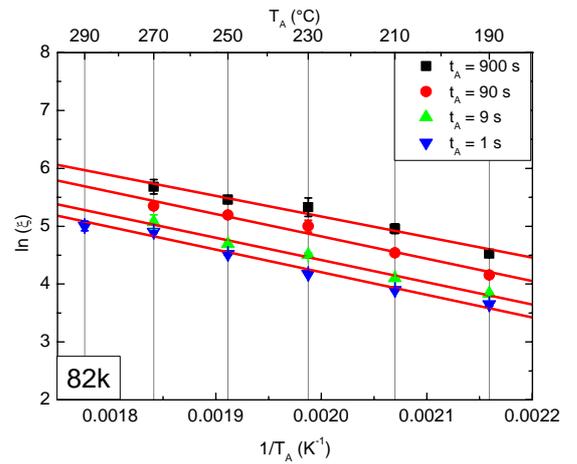

Figure 16 Arrhenius plot of $\xi$ data at different $t_A$.

## S5:      N = 987, $M_n$ = 102 kg/mol

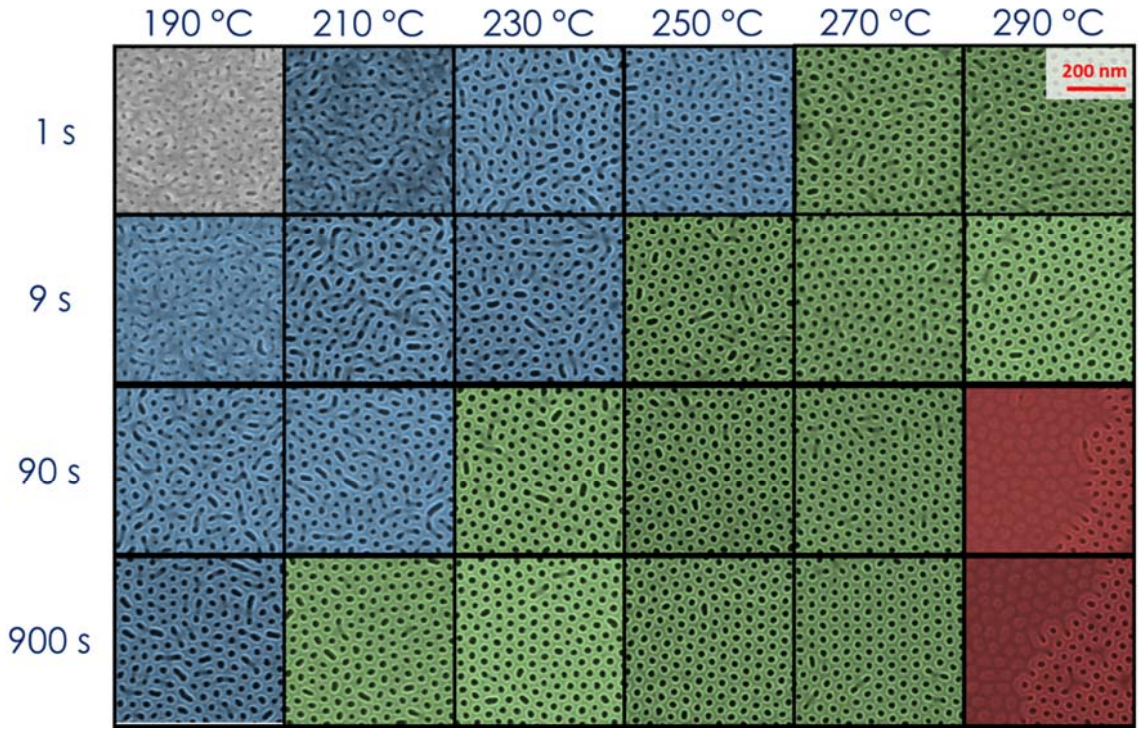

Figure 5 Plan view high-magnification SEM images of 102 kg/mol BCP thin films annealed at different temperatures $T_A$ for different $t_A$. $\xi < L_0$ gray images, $1 < \xi < 2 L_0$ blue images, $\xi > 2L_0$ green images, inhomogeneities red images.

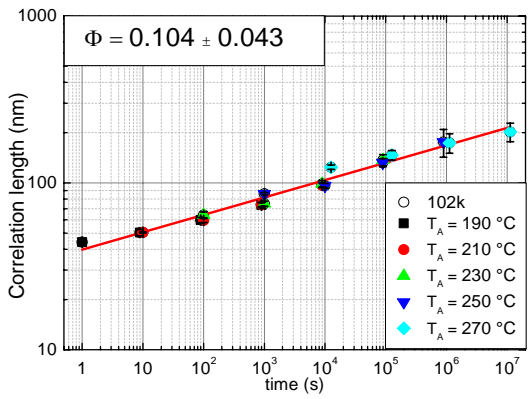

Figure 11 Master curve from time–Temperature-Superposition of the correlation length data

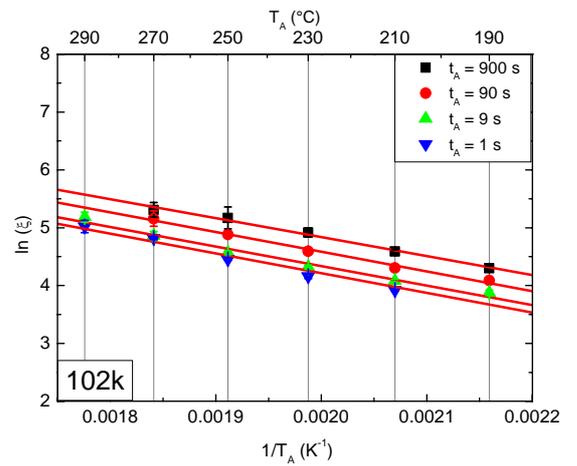

Figure 17 Arrhenius plot of $\xi$ data at different $t_A$.

## S6:       N = 1281, $M_n$ = 132 kg/mol

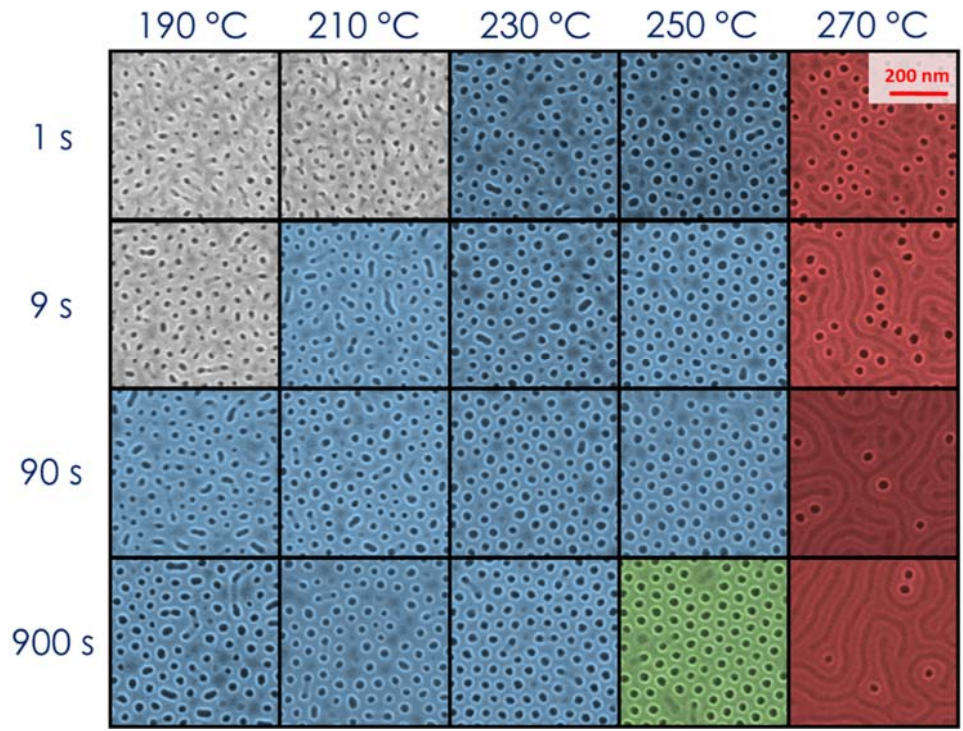

*Figure 6 Plan view high-magnification SEM images of 132 kg/mol BCP thin films annealed at different temperatures $T_A$ for different $t_A$. $\xi < L_0$ gray images, $1 < \xi < 2\,L_0$ blue images, $\xi > 2L_0$ green images, inhomogeneities red images.*

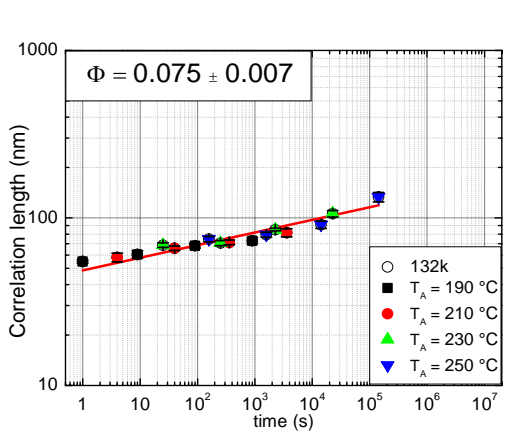

*Figure 12 Master curve from time–Temperature-Superposition of the correlation length data*

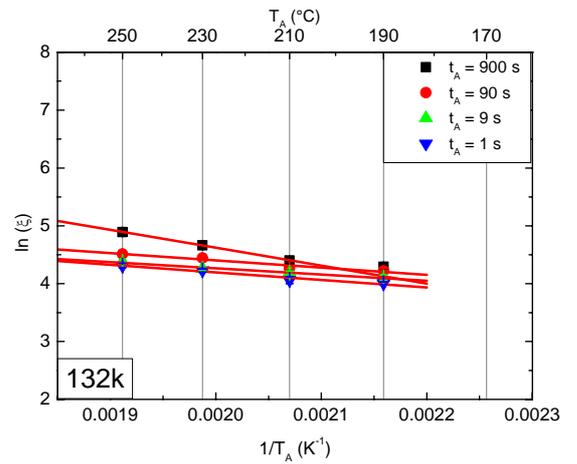

*Figure 2 Arrhenius plot of $\xi$ data at different $t_A$.*